\documentclass[12pt]{article}

\usepackage{amsfonts, amsmath, amssymb, amsthm}
\usepackage{a4}
\usepackage{graphicx}

\DeclareMathOperator{\atanh}{atanh}

\theoremstyle{definition}
\newtheorem{cnv}{Convention}

\theoremstyle{remark}
\newtheorem{remark}{Remark}
\newtheorem{case}{Case}

\author{I.G. Korepanov}
\title{Three-dimensionalizing the eight-vertex model}
\date{January 2016}

\begin{document}

\maketitle

\begin{abstract}
A simple ansatz is proposed for two-color $R$-matrix satisfying the tetrahedron equation. It generalizes, on one hand, a particular case of the eight-vertex model to three dimensions, and on another hand --- Hietarinta's permutation-type operators to their linear combinations. Each separate $R$-matrix depends on one parameter, and the tetrahedron equation holds provided the quadruple of parameters belongs to an algebraic set containing five irreducible two-dimensional components.
\end{abstract}

\section{Introductory remarks}\label{s:intro}

\subsection[Two-dimensional case: permutation-type $R$-matrices in the Baxter${}\cap{}$Felderhof model]{Two-dimensional case: permutation-type $R$-matrices in the Baxter${}\boldsymbol{\cap}{}$Felderhof model}\label{ss:BF}

Below we tend to use the notations from Hietarinta's paper~\cite{hietarinta}. Recall that Hietarinta proposed in that paper solutions to Yang--Baxter, tetrahedron, and higher simplex equations, acting as permutations on (tensor products of) basis vectors. For instance, the following simple permutation yields a solution to the constant Yang--Baxter equation (CYBE):
\[
\mathcal S(e_i\otimes e_j)=e_j\otimes e_i
\]
(compare~\cite[formula~(9)]{hietarinta}).

\begin{cnv}\label{cnv:2c}
In this paper, we consider only \emph{two-color} solutions, the colors being denoted as $0$ and~$1$. So, our indices $i,j,\ldots$ take only these two values.
\end{cnv}

\begin{cnv}\label{cnv:am}
The addition of indices mentioned in Convention~\ref{cnv:2c} is understood modulo~$2$.
\end{cnv}

There is one more interesting operator, although it does not make by itself a solution to CYBE:
\[
\mathcal T=\begin{pmatrix}0&1\\ 1&0\end{pmatrix} \otimes \begin{pmatrix}0&1\\ 1&0\end{pmatrix} \cdot \mathcal S, \qquad \mathcal T(e_i\otimes e_j)=e_{j+1}\otimes e_{i+1}.
\]
Consider, however, the following \emph{linear combination}:
\begin{equation}\label{R-YB}
\mathcal R(\lambda) = \mathcal S+\lambda \mathcal T,
\end{equation}
and the following \emph{non-constant} Yang--Baxter relation:
\begin{equation}\label{ncYB}
\mathcal R_{12}(\lambda) \mathcal S_{13}(\mu) \mathcal T_{23}(\nu) = \mathcal T_{23}(\nu) \mathcal S_{13}(\mu) \mathcal R_{12}(\lambda).
\end{equation}
One can readily see that \eqref{ncYB} holds provided
\begin{equation}\label{lmn}
\lambda-\mu+\nu-\lambda\mu\nu=0.
\end{equation}

\begin{figure}[t]
 \centering
\includegraphics[scale=0.6]{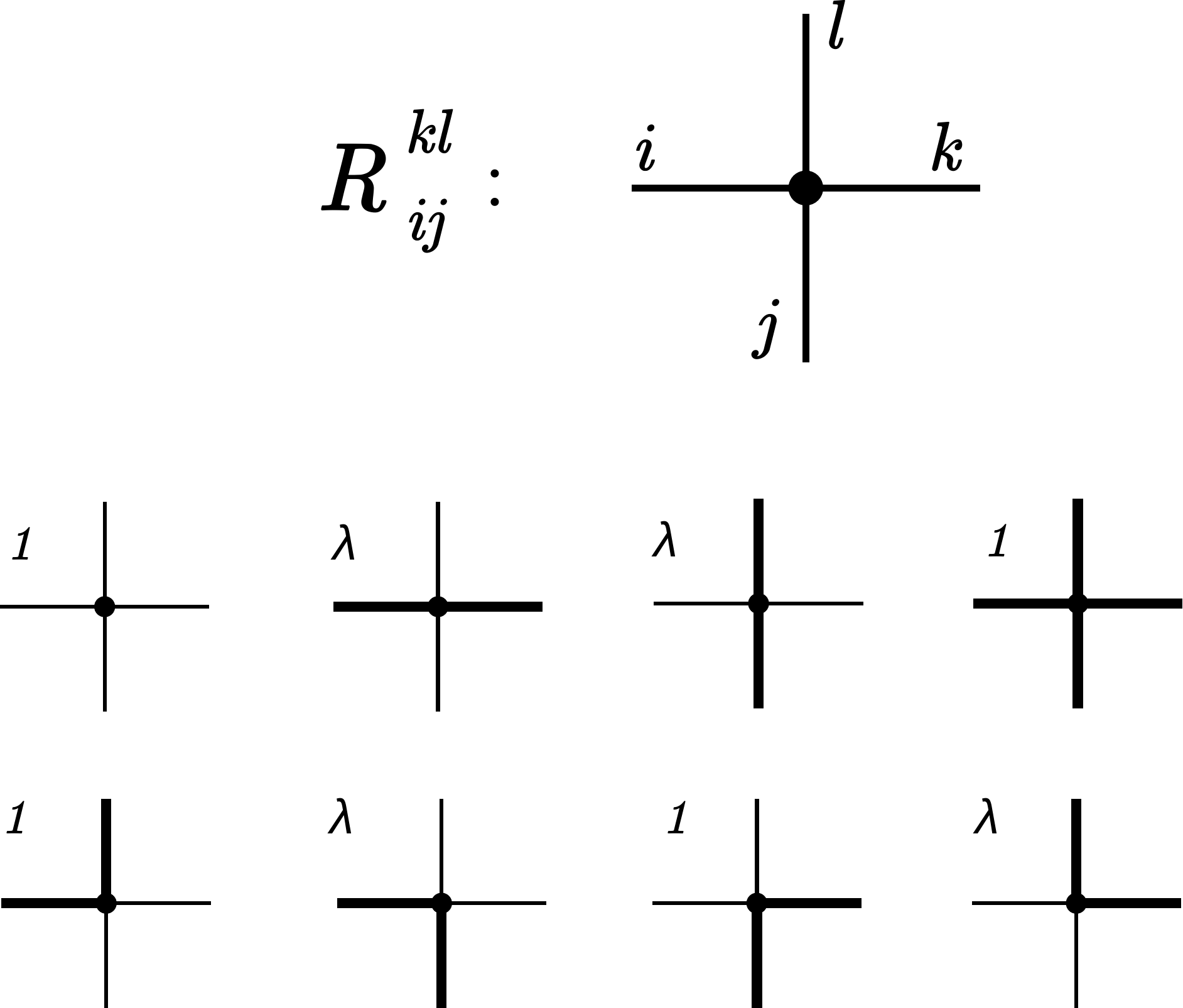}
 \caption{Matrix elements of operator~\eqref{R-YB}. Thin lines correspond to color~$0$, thick lines --- to color~$1$}
 \label{f:8}
\end{figure}

Our operator~\eqref{R-YB} is a particular case of both Baxter~\cite{baxter} and Felderhof~\cite{felderhof} eight-vertex $R$-operators. Figure~\ref{f:8}
depicts its matrix elements.

\begin{remark}
A reader with experience in various kinds of eight-vertex models will notice that our $R$-operator~\eqref{R-YB} can be represented, by taking different bases in our two-dimensional ``color'' spaces, by just a ``four-vertex'' $R$-matrix. Nevertheless, some simple further analysis shows that the corresponding lattice statistical model displays not completely trivial behavior, with a phase transition at $\lambda=0$.
\end{remark}

\begin{remark}
Also, imagine that we knew only the model with Boltzmann weights as in Figure~\ref{f:8}, satisfying equation~\eqref{ncYB} --- then this could be already a good stimulus for (re)discovering Baxter's and Felderhof's models, by substituting different weights instead of ones and lambdas.
\end{remark}

\begin{remark}
And finally, $R$-operator~\eqref{R-YB} will soon be generalized to much less trivial case of three dimensions, compare \eqref{lmn} with the formula in Case~\ref{case:3} in Section~\ref{s:lcp}.
\end{remark}

\subsection[Commuting transfer matrices from not necessarily invertible $R$-operators]{Commuting transfer matrices from not necessarily invertible $\boldsymbol{R}$-operators}\label{ss:non-inv}

Sometimes we will encounter \emph{non-invertible} $R$-operators satisfying Yang--Baxter or tetrahedron equation. A simple example is operator~\eqref{R-YB} with $\lambda=\pm 1$. We would like to remark here that such operators can still be relevant for constructing commuting transfer matrices.

\begin{figure}[t]
 \centering
\includegraphics[scale=0.5]{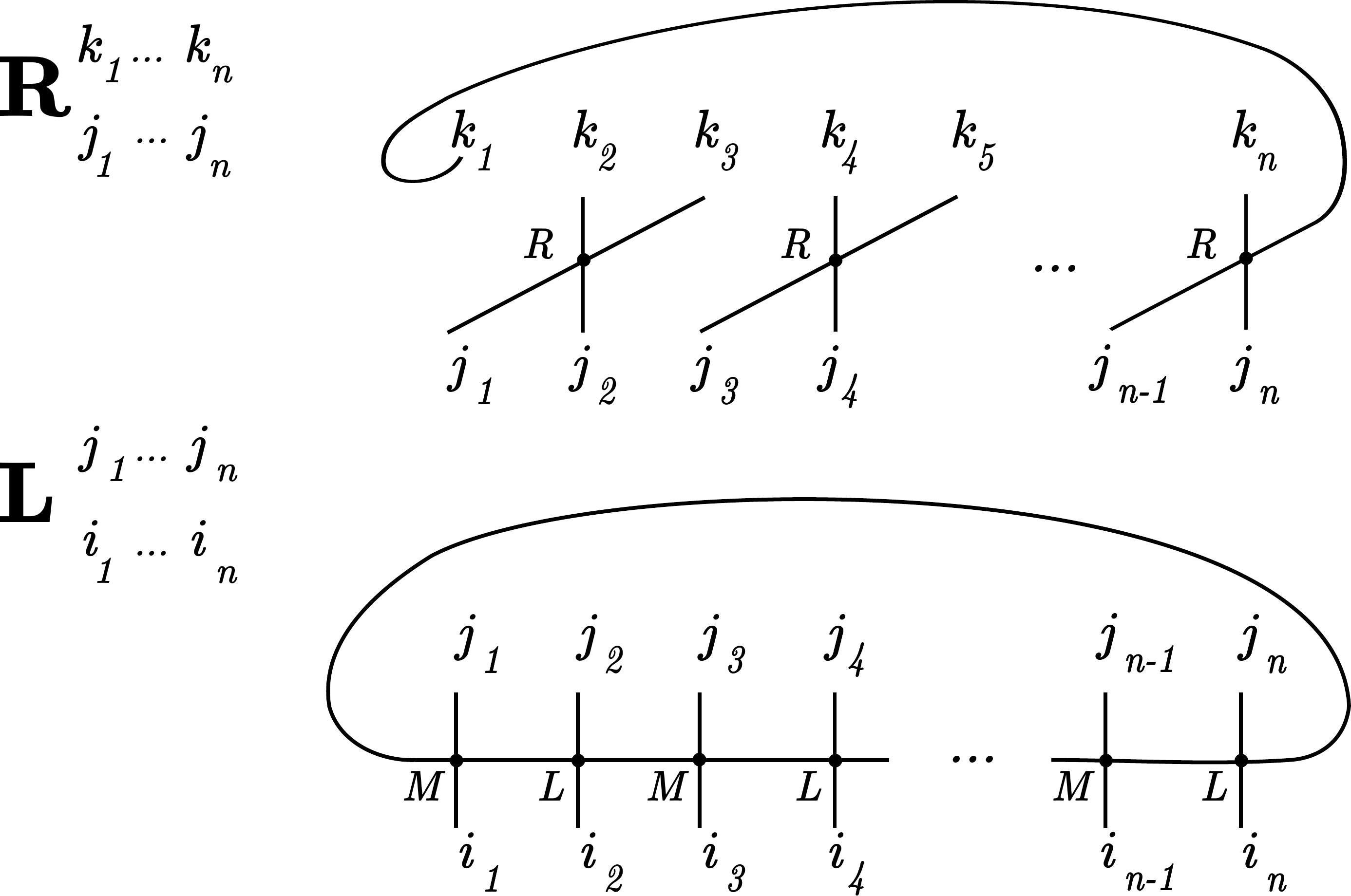}
 \caption{These transfer matrices commute if \eqref{RLM} holds.}
 \label{f:d}
\end{figure}

To explain the idea, it is enough to take the Yang--Baxter case. So, let the relation
\begin{equation}\label{RLM}
\mathcal R_{12}\mathcal L_{13}\mathcal M_{23} = \mathcal M_{23}\mathcal L_{13}\mathcal R_{12},
\end{equation}
with no requirements of invertibility for $\mathcal R$, $\mathcal L$ and~$\mathcal M$. Then, the two commuting transfer matrices $\mathbf R_{j_1\dots j_n}^{k_1\dots k_n}$ and~$\mathbf L_{i_1\dots i_n}^{j_1\dots j_n}$ are most easily presented pictorially, see Figure~\ref{f:d}.

Similar construction of commuting transfer matrices in the case of one more dimension, and tetrahedron equation instead of Yang--Baxter, can be found in paper~\cite{korepanov}: the analogue of the upper transfer matrix in Figure~\ref{f:d} consists of ``hedgehogs'', while the other transfer matrix forms a \emph{kagome} lattice, with three different types of vertices. We refer the reader to~\cite{korepanov} for details.

\subsection{Taking partial trace of Hietarinta's solution to the four-simplex equation}\label{ss:hie-4s}

Recall that Hietarinta's permutation-type $R$-operators were defined, for an $n$-simplex equation, using an $(n\times n)$-matrix~$A$ and $n$-column~$B$, usually written together as $[A|B]$, and acting \emph{linearly on indices} in the following way~\cite[formula~(10)]{hietarinta}:
\[
R_{i_1\dots i_n}^{j_1\dots j_n} = \delta_{A_1^{\alpha}i_{\alpha}+B_1}^{\,j_1} \dots \delta_{A_n^{\alpha}i_{\alpha}+B_n}^{\,j_n}.
\]
In particular, Hietarinta discovered the following solution to the \emph{four}-simplex equation~\cite[Subsection~6.21]{hietarinta}:
\begin{equation}\label{R4}
 \left[\begin{array}{cccc|c}
  1 & 1 & 1 & 1 & 0 \\ 0 & 0 & 1 & 1 & 0 \\ 0 & 1 & 0 & 1 & 0 \\ 0 & 0 & 0 & 1 & 0
 \end{array}\right].
\end{equation}

It is well-known that if there is a solution to $n$-simplex equation then a solution to $(n-1)$-simplex equation can be obtained by taking a \emph{partial trace} in one pair of indices. For instance, if $\tilde{\mathcal R}_{i_1i_2i_3i_4}^{j_1j_2j_3j_4}$ enjoys the four-simplex equation, then
\[
\mathcal R_{i_1i_2i_3}^{j_1j_2j_3} = \tilde{\mathcal R}_{i_1i_2i_3 k}^{j_1j_2j_3 k}
\]
is a solution to tetrahedron. Partial trace of~\eqref{R4} in the fourth pair of indices gives the sum
\begin{equation}\label{S+T}
\mathcal R = \mathcal S + \mathcal T
\end{equation}
of two permutation-type operators, corresponding to the following two respective matrices $[A|B]$:
\begin{equation}\label{ST}
 [A|B]_{\mathcal S} = 
 \left[\begin{array}{ccc|c}
  1 & 1 & 1 & 0 \\ 0 & 0 & 1 & 0 \\ 0 & 1 & 0 & 0
 \end{array}\right],
 \qquad [A|B]_{\mathcal T} = 
 \left[\begin{array}{ccc|c}
  1 & 1 & 1 & 1 \\ 0 & 0 & 1 & 1 \\ 0 & 1 & 0 & 1
 \end{array}\right].
\end{equation}
Note that operator~$\mathcal S$ satisfies by itself the tetrahedron equation (and can be obtained by central reflection from the second solution in~\cite[Subsection~5.4]{hietarinta}).

Operator~\eqref{S+T} is highly degenerate: of rank~$4$, while acting in an eight-dimen\-sional space. It can still be already interesting, as explained in our Subsection~\ref{ss:non-inv}. Remarkably, it can also be generalized, according to formula~\eqref{S+aT} below, while still obeying the (now non-constant) tetrahedron equation.

\section{Linear combinations of permutation-type solutions to the tetrahedron equation, and a 16-vertex model in three dimensions}\label{s:lcp}

We consider the following non-constant tetrahedron equation:
\begin{multline}\label{Ra}
\mathcal R_{123}(a_{123}) \mathcal R_{145}(a_{145}) \mathcal R_{246}(a_{246}) \mathcal R_{356}(a_{356}) \\
= \mathcal R_{356}(a_{356}) \mathcal R_{246}(a_{246}) \mathcal R_{145}(a_{145}) \mathcal R_{123}(a_{123}),
\end{multline}
where all $\mathcal R_{\alpha\beta\gamma}$ are defined according to the following ansatz:
\begin{equation}\label{S+aT}
\mathcal R_{\alpha\beta\gamma}(a_{\alpha\beta\gamma}) = \mathcal S_{\alpha\beta\gamma} + a_{\alpha\beta\gamma}\mathcal T_{\alpha\beta\gamma},
\end{equation}
and $\mathcal S_{\alpha\beta\gamma}$ and~$\mathcal T_{\alpha\beta\gamma}$ are operators $\mathcal S$ and~$\mathcal T$, defined according to~\eqref{ST} and acting in the tensor product of spaces numbered $\alpha$, $\beta$ and~$\gamma$.

\begin{figure}
 \centering
\includegraphics[scale=0.6]{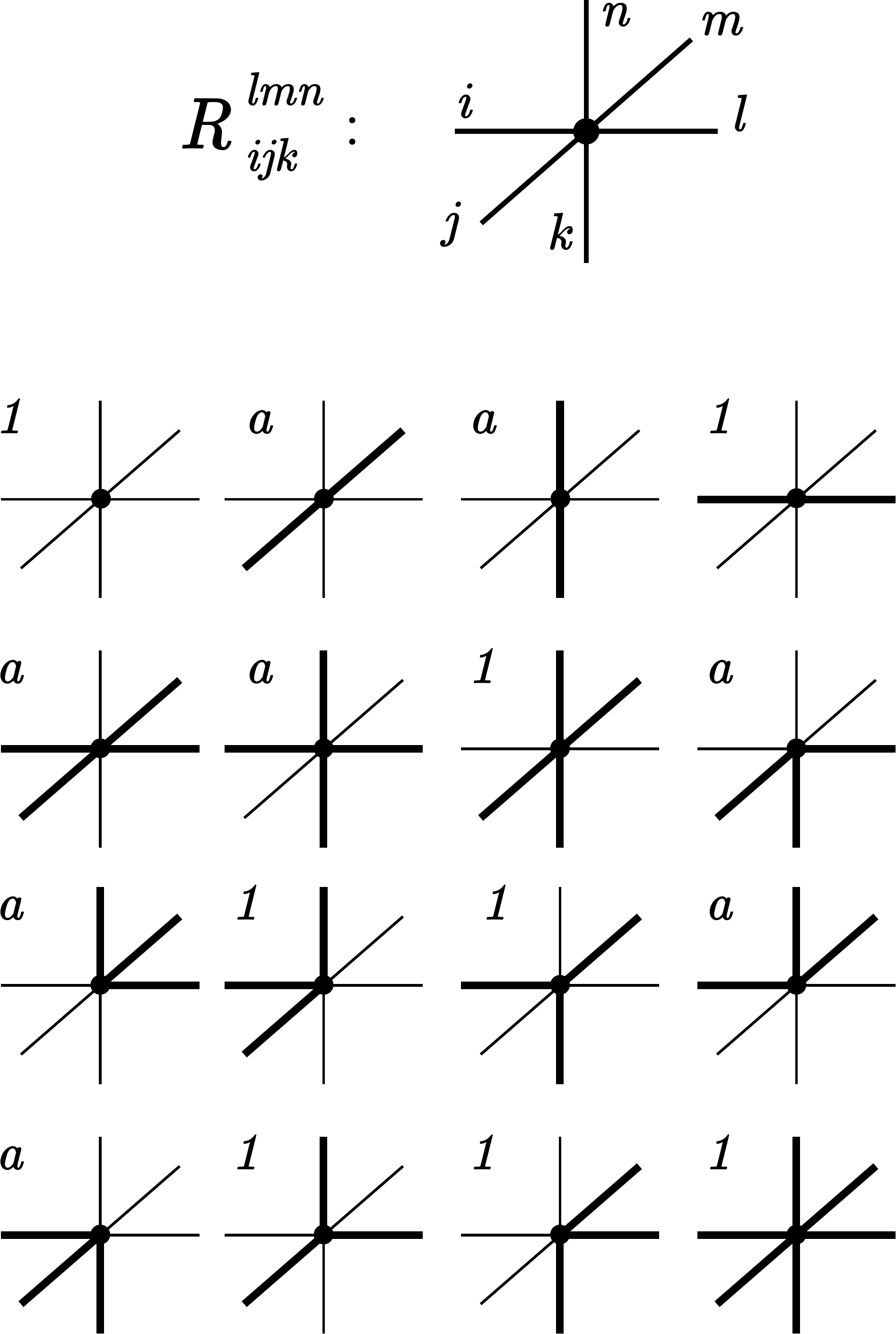}
 \caption{Matrix elements of operator~\eqref{S+aT}, with subscripts $\alpha$, $\beta$ and~$\gamma$ left out. Thin lines correspond to color~$0$, thick lines --- to color~$1$}
 \label{f:S+aT}
\end{figure}

Matrix elements of operator~\eqref{S+aT} are depicted in Figure~\ref{f:S+aT}.
Operator~\eqref{S+aT} can thus be said to correspond to a three-dimensional \emph{sixteen-vertex} model.

Equation~\eqref{Ra} imposes some restrictions on the four parameters $a_{123}$, $a_{145}$, $a_{246}$ and~$a_{356}$. Namely, a calculation using Singular\footnote{https://www.singular.uni-kl.de/} computer algebra system shows that \eqref{Ra} holds provided they belong to the algebraic variety containing five irreducible components written out below as ``Cases \ref{case:1}--\ref{case:5}''.

\begin{case}\label{case:1}
\[
a_{356}-1=0,\qquad a_{246}-1=0.
\]
\end{case}
\begin{case}\label{case:2}
\[
a_{356}+1=0,\qquad a_{246}+1=0.
\]
\end{case}
\begin{case}\label{case:3}
\[
a_{145}a_{246}a_{356}-a_{145}+a_{246}-a_{356}=0,\qquad a_{123}-1=0.
\]
\end{case}
\begin{case}\label{case:4}
\[
a_{145}a_{246}a_{356}-a_{145}-a_{246}+a_{356}=0,\qquad a_{123}+1=0.
\]
\end{case}
\begin{case}\label{case:5}
\[
a_{246}-a_{356}=0,\qquad a_{145}=0.
\]
\end{case}

\section{Discussion}

This short note shows that there exists some interesting algebra related to the 16-vertex model in Figure~\ref{f:S+aT}. It is not yet known how nontrivial this model is or/and what interesting generalizations it admits. One feature stimulating further research in this direction is that Boltzmann weights in Figure~\ref{f:S+aT} are clearly positive when $a>0$.

Note that all $R$-operators are (generically) non-degenerate in our Case~\ref{case:5}. Other cases are also of interest, for instance, the relation between parameters in Case~\ref{case:3}, if written in the additive form
\[
\atanh a_{246} = \atanh a_{145}+\atanh a_{356},
\]
resembles the usual Yang--Baxter models very much --- although more effort could be required to cope with the degeneracy of~$\mathcal R_{123}$.

\end{document}